# Nonequilibrium assembly of Lennard-Jones particles on a sphere

*Ivan Yu. Golushko, Olga V. Konevtsova, Daria S. Roshal, Sergei B. Rochal.\**

Physics Faculty, Southern Federal University, Rostov-on-Don 344090, Russia.

ABSTRACT: Studying physical mechanisms and common geometric principles underlying known spherical packings is crucial for rational design of synthetic nanocontainers. Here we model the growth of small spherical shells containing $n \leq 72$ identical particles that have their own curvature and interact with each other via the Lennard-Jones potential. The shell assembly is assumed to be nonequilibrium and sequential: at each step, a new particle is attached to the most energetically favorable position, after which the system relaxes. Along with well-known structures of the smallest icosahedral viral protein shells, the proposed mechanism generates a wide range of shells exhibiting square-triangular surface order. Most of such shells are the models of synthetic or natural protein complexes that have octahedral or tetrahedral symmetries and perform various functions. We compare the obtained structures with those resulting from the equilibrium assembly and corresponding to global energy minima. Also, we consider the temperature-dependent stochastic assembly and use the double-minimum Lennard-Jones-Gauss potential to mimic anisotropic particle interactions.

A plethora of examples of two-dimensional packings of identical or nearly identical structural units (SUs) can be found in both living and non-living nature.[1] If interactions between SUs are isotropic and can be described by a single-minimum potential then in the low-temperature limit, planar systems tend to form a dense packing, in which the SUs are located at the vertices of a simple triangular lattice.[2] The arrangement of spherical structures is inherently more complex since attempting to wrap the triangular lattice around a sphere leads to the formation of topological defects. According to Euler's theorem, even in the most ordered packing, there should be 12 point disclination defects. When bypassing a closed loop surrounding the defect, the lattice translation will not coincide with itself, but will rotate by 60 degrees, which corresponds to a 60-degree sector cut out of the planar lattice prior to its mapping onto the spherical surface. These defects occupy positions coinciding with the vertices of a regular icosahedron inscribed in a sphere. In general, types of defects and their positions can vary resulting in more defective structures, as occurs in a number of systems with extended topological defects,[3-6] including colloidal crystals.[7-8]

The most perfect triangular order is observed in the protein shells of icosahedral viruses (capsids), formed by complexes of 5 or 6 proteins, the so-called capsomeres.[9-10] The interplay of protein geometry, repulsive electrostatic and attractive hydrophobic interactions define the equilibrium shape and properties of these shells. Despite this complexity, even strongly idealized models can reproduce arrangement of capsomeres in these capsids, highlighting the key role of the balance between the surface curvature and in-plain interactions of structural units in the system ordering.[11-15] When modeling the self-assembly of isotropic disks retained on the spherical surface, several simple models[11-12] produce structures of the four smallest icosahedral shells, consisting of 12, 32, 42, and 72 capsomeres.

For example, the authors of the work[11] used a phenomenological Hamiltonian that depended on the angles between the normals of adjacent disks, their valence and the coverage density of the spherical surface. Using the model, they obtained a phase diagram of the equilibrium self-assembly and showed that structures consisting of 12, 24, 32, 48, and 72 identical disks correspond to local minima of energy per particle. The follow-up study[12] by the same group used a simpler model where packings were modeled as sets of particles retained on the spherical surface and interacting via the Lennard-Jones (LJ) potential. This study yielded similar results demonstrating that the addition of 12 smaller particles lowers the energies of icosahedral packings making them global minima for systems with 12, 32, 42 and 72 structural units. In the subsequent works, the model was further developed by varying number of smaller disks involved in the assembly[12,16] and replacing the LJ potential with pure repulsion.[16]

It should be noted that the packings found in nature do not necessarily represent ground states of the corresponding systems, and one of the key factors is the assembly pathway and the extent to which the system can explore the available configurational space. The processes of equilibrium and nonequilibrium assembly of spherical shells were compared in refs. 13-15, where capsids were considered as discrete elastic shells constructed from triangles. The models[13-15] along with the in-plane deformation energy of the triangles also considered a bending energy that depended on the angles between the normals of adjacent triangles. During the simulation of the nonequilibrium assembly, at each step, a new particle was added at the boundary of the growing shell according to certain geometric rules, after which the system was relaxed.[13-15] Interestingly, the phase diagrams for the cases of equilibrium and nonequilibrium assembly pathways contained many identical spherical structures with similar regions of existence in the parameter space; however, some shells were exclusive to a certain assembly type.

In this paper, we propose a more straightforward model of the nonequilibrium assembly, in which particles are retained on the spherical surface and interact via the LJ potential. At each assembly step a single particle is attached at the most energetically favorable position. The assembly stops when adding new particles is no longer energetically favorable. This algorithm corresponds to the situation where the ratio between the interface energy of the growing shell and the interaction strength between the SUs allows them to explore multiple attachment positions before becoming a part of the shell.[17] With the only parameter being the ratio of the equilibrium distance between particles to the sphere radius, we are able to perform a detailed analysis of the self-assembly of small (n≤72) shells in which the influence of curvature is most prominent. Along with the well-known icosahedral structures, the model generates highly symmetric shells with $T$, $O$, $O_h$ and $D_{5h}$ symmetries corresponding to experimentally observed synthetic and natural protein shells.

MODEL

When constructing the model, we consider SUs as particles retained on the spherical surface of a unit radius $R = 1$. The interaction between particles is described by the Lennard-Jones potential

$$V(d) = \epsilon \left[ \left(\frac{\sigma}{d}\right)^{12} - 2\left(\frac{\sigma}{d}\right)^6 \right], \tag{1}$$

where $d$ is a distance between a pair of particles, $\sigma$ is its equilibrium value and $V(\sigma) = -\epsilon$ is a depth of the energy minimum. Since $R = 1$ and the ratio $\sigma/R$ equals $\sigma$, the latter variable is the only parameter of the developed model.

We assume that the shell grows in discrete steps, and at each step, a single particle is attached in position **r** on the spherical surface, corresponding to the maximum of the binding energy absolute value $|E_b(\mathbf{r})|$. This is essentially an energy of pair interactions of the $(n + 1)^{\text{th}}$ particle with the already attached $n$ particles:

$$E_b(\mathbf{r}) = \sum_{i=1}^{n} V(|\mathbf{r} - \mathbf{r}_n|), \qquad (2)$$

After the attachment, the energy of the spherical cluster is minimized with respect to coordinates of all its particles. For this purpose, we use a gradient descent method with the imposed condition that the particles are retained on the spherical surface. When there are no positions left at which the potential (2) is negative, the assembly is stopped resulting in a shell of $n$ particles with the energy per particle being equal to $\mu = E_{tot}/n$ (where $E_{tot}$ is the total energy of the shell). To speed up the assembly simulations, the point with the minimum binding energy (2) is selected from an array of trial points located on the sphere surface. Specifically, 5000 approximately equidistant points are placed around each already attached particle in a region representing a spherical cap with a radius of $1.5\sigma$. The most energetically favorable position is then selected from this set.

Obviously, during the considered nonequilibrium shell assembly, at each step the energy landscape corresponding to the attachment energy of the next particle depends on both the position of the already attached particles and the value of $\sigma$. Nevertheless, since the attachment of new particles does not occur randomly, but at the point with the minimum binding energy (2), structurally similar shells with the same assembly pathway and $n$ appear in a certain finite interval $\Delta\sigma$. Naturally, the energies of such shells gradually change with $\sigma$. The above-mentioned assembly features are visible in Figure 1a, which shows $\mu(\sigma)$ for $\sigma$ ranging from 0.85 to 1.18. Note that structures with certain numbers of particles never assemble regardless of $\sigma$, whereas in some cases, several structurally distinct shells with the same *n* may appear. Therefore, to construct the graphs presented in Figure 1b,c, from the sets of shells with the same number of particles, we select a structure corresponding to the minimum value of $\mu$; we denote this value as $\mu_{min}$.

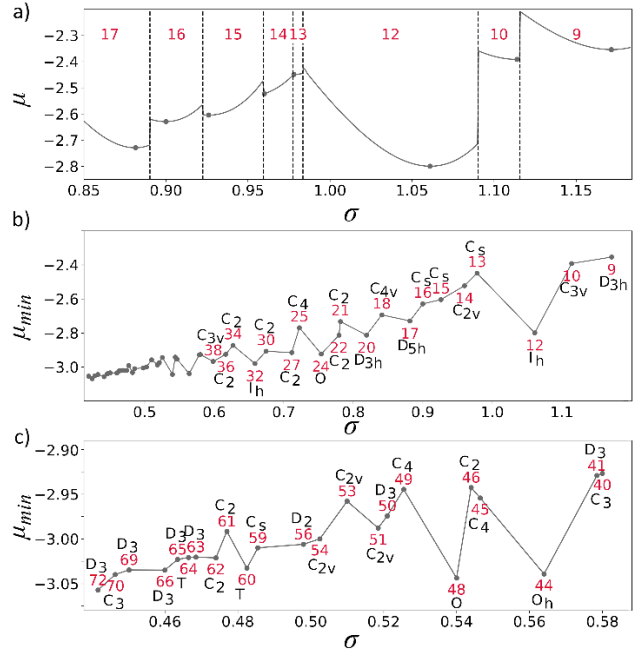

**Figure 1**. Energy spectrum of the structures resulting from nonequilibrium assembly on the spherical surface. (a) Energy per particle $\mu$ as a function of the equilibrium distance $\sigma$ in the Lennard-Jones potential. The dotted lines are the boundaries of the phases existing in the shown region. (b) Energy per particle $\mu_{min}$ for the most energetically favorable structures as a function of $\sigma$. (c) Enlarged fragment of the graph (b). In all graphs, the number of particles in the shell is shown in red, and structure symmetry is shown in black.

Note that with the increase of number of particles, the number of structurally different shells with the same *n* increases, while the energy difference between these structures decreases. Moreover, in certain intervals, a decrease in $\sigma$, instead of an increase in *n*, may lead to its decrease. For example, this occurs in the interval $0.4333 < \sigma < 0.4481$ at the beginning of which, structures with *n*=72 assemble, followed by regions with *n*=70 and then again with *n*=72.

Nevertheless, if we consider shells containing no more than 72 particles then this is the only interval with such complex behavior, and the model described above can be analyzed in sufficient detail. When searching for structures in the parameter space, we used step $\delta\sigma = 10^{-4}$, decreasing it by a factor of 2, 3 and 5 in regions with complex behavior of function $\mu(\sigma)$. We made sure that a further step decrease does not qualitatively change the behavior of the function and no new phases appear (structures with different *n* values and/or a fundamentally different arrangement of particles).

Interestingly, all the assembled minimum-energy structures have nontrivial symmetry in the considered $\sigma$ range. Out of 43 structures assembling within the proposed framework 7 have high symmetry (cubic and icosahedral symmetry), 18 have intermediate symmetry (with only one *m*-fold symmetry axis, where *m*>2), and 18 have low symmetry. Importantly all the structures with high symmetry coincide with the lowest energy states found in works,[18-19] among the structures with intermediate symmetry, there are 11 such structures, and among the low symmetry structures, there are 8 such structures.

To randomize the assembly process, we introduced effective thermal energy to the model. Previously, we attached a particle to the most energetically favorable position. Now, we first find positions corresponding to local minima of the attachment energy, and then select one of them using the Boltzmann distribution. The local minima are found as positions, which do not have neighboring trial positions, providing greater gain in binding energy; the trial positions are found as described above and considered to be neighboring within a distance of $\sigma/2$.

Let at the current step there are *q* local minima where the particle can attach. Then we find the probability $p_j$ of filling the *j*-th position as

$$p_j = \exp\left(-\frac{E_j - F}{T}\right), \tag{3}$$

where $E_j$ is the corresponding binding energy, the chemical potential $F$ is determined by the condition $\sum p_j = 1$, and $T$ is the normalized temperature. To select the attachment position we use the interval divided into segments proportional to the probabilities (3) and then generate a random number within the interval. At $T \to 0$ the growth algorithm is reduced to the previously discussed deterministic one, whereas at $T \to \infty$, the choice of local minima for attachment is completely random.

Using this algorithm, we modelled stochastic assembly for $\sigma$ values corresponding to the 26 structures, previously obtained through the deterministic assembly and coinciding with lowest energy shells found in refs. 18-19. For each of 26 $\sigma$ values we performed 60 simulations at three different temperatures $T = 0.1\epsilon$, $\epsilon$, $10\epsilon$ (20 simulations per $T$ value). At all temperatures considered, 6 out of 7 highly symmetric structures perfectly assembled in all simulations; the remaining tetrahedral structure consisting of 60 particles successfully assembled in at least half of the simulation runs.

In the case of intermediate symmetry, some structures perfectly assembled at all temperatures considered, whereas others exhibited self-assembly with a high percentage of defective shells. Except for the shells consisting of 40 and 41 particles ($C_3$ and $D_3$ symmetries), structures with medium symmetry and number of particles $N \leq 50$ successfully assembled in 90% of cases. For structures with $N = 40$ and 41 increase in temperature can decrease yield to ~50%. Assembly of structures with $N > 50$ is less efficient on average; however, for all cases temperatures considered, the portion of successfully assembled shells exceeded 10%.

Assembly of low-symmetry structures demonstrates a similar trend, with shells of 13, 14, 15, and 16 particles perfectly assembling at all temperatures considered. For structures with $N > 50$, the introduction of temperature substantially reduces the assembly efficiency; nevertheless, the proportion of successfully assembled shells does not drop below 20% even at $T = 10\epsilon$, which essentially corresponds to random choice of local minima for particle attachment.

Thus, nonequilibrium assembly in a system with a relatively small number of particles often leads to structures corresponding to global energy minima; the smaller the number of particles in the system, the greater the probability of obtaining such a structure. This is consistent with the findings of refs. 13–15, which used different models. Furthermore, within the framework of our model, we showed that nonequilibrium assembly of high symmetry structures that correspond to global energy minima is extremely efficient even at high temperatures. We expect that other model systems will exhibit analogous behavior.

RESULTS

**Shells associated with icosahedral viral capsids**

The most symmetric shells, possessing all the symmetry elements of the icosahedron, assemble only for $n$=12 and 32. They correspond to the classical Caspar-Klug (CK) viral capsids[10] with triangulation numbers T=1 and T=3, with each particle representing a center of a protein capsomere (a pentamer if located at the five-fold axis, and a hexamer otherwise). These structures are ubiquitous in nature (e.g., capsids of the Satellite tobacco mosaic (1A34) and Tobacco Necrosis (1C8N) viruses, respectively) and can be reproduced by several models with both equilibrium and nonequilibrium assembly,[13-15] moreover they are also solutions of the Thomson problem.[20] It should be noted that icosahedral viral capsids with T=1 and T=3 are more common in nature than envelopes with other T numbers. A total of 44 different T=1 viral capsids belonging to 20 viral families and 37 T=3 viral capsids from 17 families can be identified from the envelopes represented in the databases. [21-23] Notably, almost all T=1 and T=3 viral capsids belong to non-enveloped viruses with their shell lacking lipid layer and consisting exclusively of proteins.

The next two icosahedral structures allowed by the CK theory have T=4 and T=7 triangulation numbers and consist of 42 and 72 capsomers, respectively. Both of them fail to assemble within the proposed framework and do not correspond to energy minima found in works.[18-19] In our model, the 42-particle structures do not assemble at all, whereas the 72-particle structure has a lower $D_3$ symmetry. According to the authors of ref. 11, these structures can be obtained by equilibrium assembly using two types of particles with different effective radii. Since these structures are also characteristic of many viral capsids, we decided to analyze their assembly in more detail within the framework of our approach.

Following the proposed nonequilibrium assembly algorithm, we attempted to obtain structures with $n$=42 using two types of particles with radii $R_1$ and $R_2$ and the following pair interaction potential $V(d)$:

$$V(d) = \epsilon \left[ \left( \frac{R_i + R_j}{d} \right)^{12} - 2 \left( \frac{R_i + R_j}{d} \right)^6 \right]. \tag{4}$$

As is easy to verify, for such a potential the equilibrium distance between particles equals $R_i + R_j$, thus at each assembly step both the position and the type of the attached particle could be selected. However, the choice of the particle type may be ambiguous due to identical energies of equilateral and isosceles triangles formation at certain assembly steps meaning that different particle configurations can have the same energy.

To eliminate the ambiguity and keep assembly trajectories deterministic, we associate additional constant energy (not depending on the particle coordinates) with each particle type $\mu_1$ and $\mu_2$. The difference between these energies makes attachment of certain particle type more probable and acts as an additional parameter controlling the assembly. By varying $\mu_1$-$\mu_2$ in the region $R_1 \approx$ 0.58, $R_2 < R_1$, we found several structures with $n$=42, which were absent in the model with identical particles. Unfortunately, we were unable to obtain 42-particle structures with $I_h$ symmetry within the proposed framework.

The T=4 shells emerge in more complex models. In ref. 13 capsids are modeled as elastic shells described by the discrete energy proposed earlier in ref. 24, and their non-equilibrium assembly involves formation of irreversible bonds between particles. In the ref. 25 the T=4 capsid resulted from molecular dynamics simulations of triangular subunits assembling around a flexible genome.

It should be noted that most viruses with T=4 shells are surrounded by an outer lipid envelope. This feature is characteristic of viruses from two of the three existing T=4 virus families: Togaviridae and Hepadnaviridae.[22] The most studied of these viruses, Hepatitis B virus, exhibits so-called dimorphism,[26] which is the ability to form both T=3 and T=4 shells simultaneously, albeit with different proportions of the two capsid species. In living cells, the virus primarily assembles on lipid membrane forming T=4 shells. In vitro, its proteins self-assemble into empty T=3 and T=4 structures in a ratio of approximately 95:5.[26-27] Thus, outer lipid layer likely influences the self-assembly process of these viruses and, along with anisotropic interactions between larger hexamers and smaller pentamers can stabilize the T=4 capsids. We believe these two factors are responsible for the discrepancies between predictions of our relatively simple model and experimentally observed structures.

Similarly, in the case of T=7 capsids, additional proteins that are often attached to the shell from the inside might be an additional factor stabilizing icosahedral symmetry. In our model, lacking these assembly features, the lowest energy 72-particle structure formed at $\sigma = 0.44$, has a lower $D_3$ symmetry and represents a slightly deformed T=7 icosahedral shell, in which twelve pentavalent particles are distributed over two symmetrically non-equivalent orbits; see Figure 2a.

Considering a small increase in σ value up to $\sigma = 0.448$, we note that another highly ordered phase with n=72 and $D_{5h}$ symmetry assembles within our model; see Figure 2c. Despite its higher symmetry compared to $D_3$, the phase has a higher energy and is therefore not shown in Figure 1. More complex models[13-14,18] also predict existence of the $D_{5h}$ shell, moreover in these models it is the most energetically favorable structure for n=72.

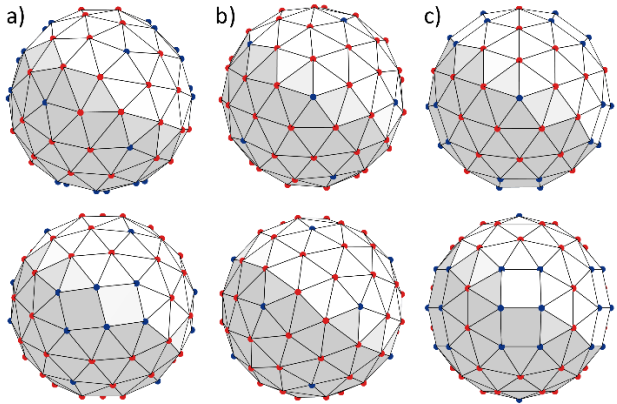

**Figure 2**. Model structures consisting of 72 particles. (a-b) deformed ($D_3$ symmetry) and ideal ($I_h$ symmetry) T=7 shells. (c) Shell with $D_{5h}$ symmetry. In the top row, the highest-order axis of each structure is perpendicular to the figure plane; in the bottom row the same applies to two-fold axis. The blue and red nodes have 5 and 6 nearest neighbors, respectively.

By sufficiently increasing the value of σ and minimizing the energy of all pair interactions (1) in the already assembled structures with n=72, the latter can eventually be transformed into an ideal T=7 icosahedral shell; see Figure 2b. A detailed study of this phenomenon can be performed within the framework of catastrophe theory (see for example ref. 28) and is beyond the scope of the present paper. Nevertheless, we emphasize that the T=7 icosahedral shell cannot be a global minimum of energy (2) if σ is close to the diameter of the spherical cap in non-overlapping dense packing.

In the context of the polymorphism observed, it is interesting to note that at the initial stages of self-assembly of real T=7 shells, a so-called procapsid is formed (e.g., in bacteriophages HK97 and P22) which later transforms into a stable infectious virion by passing through several intermediate states.

**Shells with square-triangular order**

Although icosahedral viral capsids described by the classic CK theory[10] encompass the majority of natural protein shells, there are many examples of structures, in which proteins form a square-triangular (S-T) order.[29] Such shells usually consist of less than 60 proteins (the size of the smallest icosahedral viral capsid), have $O$ or $T$ symmetry and perform a wide range of functions: they store toxic iron,[30-31] vitamin E,[32] carry out protein folding and degradation,[33-34] catalyse chemical reactions in living systems.[35-36]

The minimalistic theoretical framework we proposed here describes self-assembly of several structures with the S-T order. These structures with n=24, 44, 48, 60, 64 and symmetries $O$, $O_h$, $O$, $T$, $T$, respectively, are shown in Figure 3a-e.

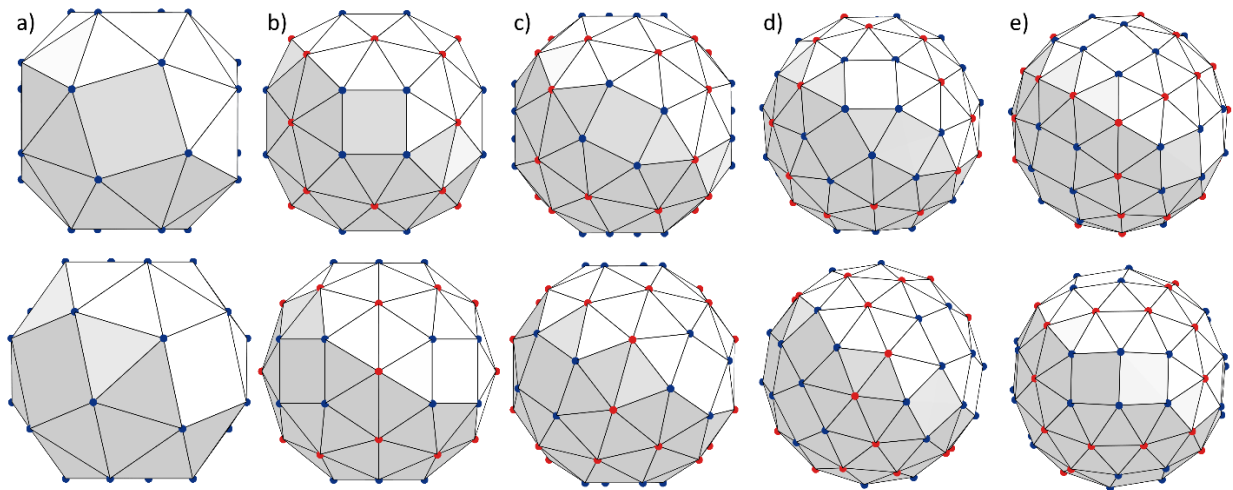

**Figure 3**. Highly symmetric structures with the square-triangular order resulting from a nonequilibrium assembly of particles interacting via the Lennard-Jones potential. (a-e) Structures

with n=24, 44, 48, 60, 64 particles, symmetries $O, O_h, O, T, T$ and assembling at $\sigma =0.754, 0.564,$ 0.54, 0.4825, 0.4665, accordingly. In the top row, the highest-order axis of each structure is perpendicular to the figure plane; in the bottom row, the same applies to two-fold axis. The blue and red nodes have 5 and 6 nearest neighbors, respectively.

The chiral structure shown in Figure 3a is a snub cube (the well-known Archimedean solid). Within the framework of our nonequilibrium assembly model, this polyhedron with $O$ symmetry and n=24 vertices has a minimum energy at $\sigma = 0.77$. Our analysis of structures deposited Protein Data Bank (PDB)[21] shows that snub cube geometry is quite common among protein shells with cubic symmetries. For example, a similar structural organization is exhibited by the nanoparticle self-assembled from alpha-Tocopherol Transfer Protein,[37] used by cells for vitamin E transport, shown in Figure 4a. The mass centers (MCs) of the proteins in this shell almost exactly coincide with the nodes of the structure produced by our model (see the bottom row of Figure 4a). The snub cube geometry is also characteristic of the well-known enzyme imidazoleglycerol phosphate dehydratase, involved in histidine metabolism,[38] and ferritin cages, which serve as the main intracellular iron depot.[30]

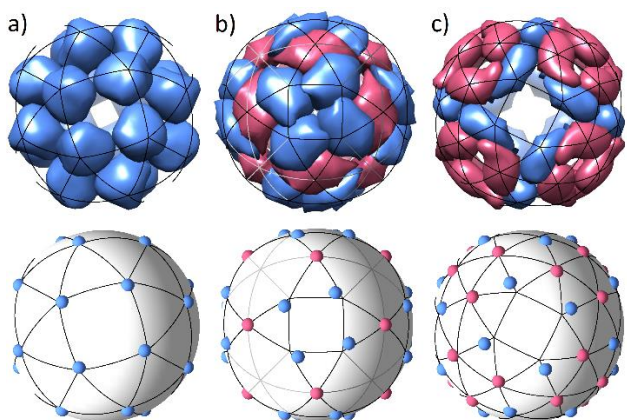

**Figure 4**. Comparison of real protein shells and model structures. Top row shows the shells (PDB IDs: 5MUE, 2VTU, and 4PO5) with the corresponding superimposed spherical lattices, the nodes

of which coincide with the vertices of the polyhedra shown in Figure 3a-c. The bottom row shows the positions of the mass centers of the proteins, projected onto the unit sphere for comparison.

Importantly, the positions of the particles in the model structure can correspond not only to individual proteins but also to their complexes. For example, protein pentamers in the case of Aquifex aeolicus lumazine synthase (AaLs) cages 5MQ3, 7A4F, 9G3O and eleven-membered ring proteins in the case of cages 6RVV, 6RVW self-assemble into structures where the mass centers of these complex SUs form a snub-cube. In a 44-particle packing, some of the particles are located on the two- and three-fold axes (see Figure 4b), and these positions can be occupied only by SUs with matching symmetry; otherwise, the positions remain vacant, as in small icosahedral viruses, where asymmetric proteins cannot occupy positions on the axes of the icosahedron.[39] For example, in the synthetic octahedral structure of 2VTU, assembling from 36 dimers, formed by two bacteriophage MS2 coat proteins each,[40] the positions of MCs of the dimers almost perfectly coincide with the nodes of the packing we obtained (see Figure 4b). The nodes on the two-fold symmetry axes are occupied by dimers, while those on the three-fold symmetry axes remain vacant. Ferritin mutant 6IPO, consisting of 48 proteins, 24 of which form dimers on the two-fold symmetry axes,[41] can also be described within the framework of the S-T spherical lattice shown in Figure 4b.

Structure 4PO5 (Figure 4c) consists of 48 individual allophycocyanin B (AP-B) proteins. AP-B is one of the two terminal emitters in phycobilisomes, the unique light- harvesting complexes of cyanobacteria and red algae.[42] The AP-B protein MCs in crystal structure 5PO5 coincide well with the vertices of the model shell with n=48; (see Figure 4c). Note that in the model structure, the triangles lying around the two-fold diagonal axes of the $O$ symmetry group are deformed, which brings their positions closer to those of the protein MCs in the real structure.

An interesting feature shared by all of the packings in Figure 4 is that they not only correspond to the global energy minima, but also their assembly in terms of our model is very robust with respect to temperature; see last paragraphs of MODEL section. This fact may partially explain why it's exactly these structures that are observed in nature. In this context, we note that the shell assembly shown in Fig. 3e is also highly probable, and we hope that real protein shells with a similar structure will be discovered soon.

DISCUSSION

The structures we obtained here confirm that the emergence of the S-T order in spherical packings assembled from a small number of SUs is an intrinsic property of the system geometry that does not necessarily require anisotropy of interparticle interaction and appears at specific ratios between the size of SUs and their curvature radius. While in the planar case, the emergence of square-triangular order can be explained by the competition between the tendencies toward close packing and the maximization of the entropic contribution to free energy,[43] in the model structures obtained above, the appearance of square tiles is solely due to the spherical topology. All the packings contain 24 topological defects, corresponding to vertices common to four triangles and one square. Along with the structures shown in Figure 3, the S-T order is also exhibited by the structures consisting of 72 particles shown in Figure 2a,c (in the former, the squares are deformed). Nevertheless, there are more than a dozen different experimentally observed types of S-T shells,[29] which clearly exceeds the number of shells that can be assembled in the proposed minimalistic model.

To expand the number of such model shells, one can directly take into account anisotropic interactions between proteins. If we remain within a framework that uses point-like SUs, in order to model anisotropic interactions, the corresponding energy should promote the formation of certain angles in triplets of nearest neighbors by considering three-particle interactions. An alternative and more convenient method (in terms of the energy minimization efficiency) is to use the two-minimum pairwise Lennard-Jones-Gauss (LJG) potential:[44]

$$V(d) = \epsilon \left[ \left(\frac{\sigma_1}{d}\right)^{12} - 2\left(\frac{\sigma_1}{d}\right)^{6} + \zeta \exp\left(-\frac{(d-\sigma_2)^2}{2D}\right) \right], \quad (5)$$

in which $\sigma_1$ and $\sigma_2$ determine the positions of the minima, and $\zeta$ controls the ratio between their depths. Potential (5) promotes the formation of certain three-particle configurations and thus mimics the anisotropic interaction. In particular, in ref. 44 the authors used molecular dynamics to demonstrate that if $\sigma_2/\sigma_1 \approx 1.9$, a planar defective S-T order with three-particle configurations with an angle of 150° is formed. In this case, for the corresponding isosceles triangle (one side of which belongs to the square, and the other belongs to the adjacent regular triangle), the ratio of its base length to that of lateral sides is $\sigma_2/\sigma_1 = \sqrt{2+\sqrt{3}}$; on a spherical surface this ratio decreases with curvature.

The article[43] motivated us to examine self-assembly of particles interacting via potential (5) within the proposed nonequilibrium approach. Since the potential depends on four independent parameters, a complete study of the parameter space is quite resource-intensive. Therefore, we first checked the stability of pre-assembled shells with the S-T order[29] observed in nature. If $\sigma_1$ and $\sigma_2$ are calculated from the structural models of these shells[29] and coefficients $\zeta$ and D are adjusted so that both minima of potential (5) are well resolved and have approximately the same depth and width, then all the model shells with n≤60 considered in ref. 29 (see Figures. 1, 6, 7, 8 therein) turn out to be stable with respect to small tangent displacements of the vertices (at least $0.01\sigma_1$).

Next, by varying $\sigma_1$ and $\sigma_2$ in the vicinity of the initially obtained equilibrium values, we performed multiple nonequilibrium self-assembly runs. The only two structures we were able to reproduce are shown in the left column of Figure 5 and correspond to two sets of parameters ($\sigma_1 = 0.716$, $\sigma_2 = 1.0141$, $\zeta = -0.768$), and ($\sigma_1 = 0.661$, $\sigma_2 = 1.206$, $\zeta = -0.9465$); in both cases D=0.005. The fact that in the considered approximation, not all stable shells self-assemble can probably be explained by two main factors. First, the LJG potential is based on the averaged distances $\sigma_1$ and $\sigma_2$ to the first and second neighbors and therefore cannot fully substitute 3-particle interactions favoring formation of certain angles between the corresponding pairs of bonds. In spherical structures, the spread of distances to the second neighbors can be significant, and the broadening of the second minimum in potential (5) can additionally favor irrelevant (asymmetric) spherical configurations. Therefore, a precise consideration of three-particle interactions could lead to the assembly of a larger number of the shells observed in nature.

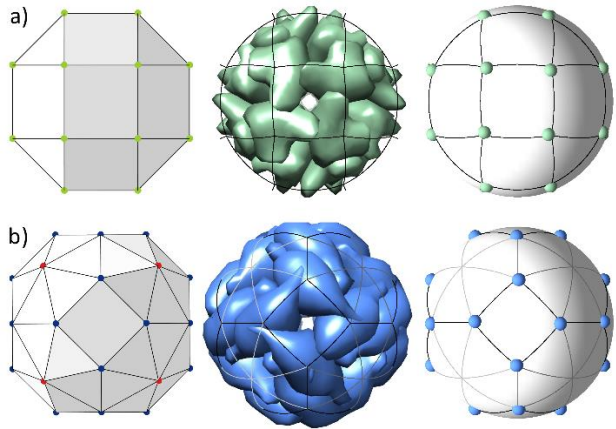

**Figure 5**. Structures resulting from a nonequilibrium assembly with the Lennard-Jones-Gaussian potential and real protein shells. Left column shows model structures with n=24 and 32 particles and $O_h$ symmetry. Middle column demonstrates shells 3RA0, 9NV4 with corresponding superimposed spherical lattices, the nodes of which coincide with the vertices of the polyhedra shown in the left column. Right column shows the positions of the mass centers of the proteins projected onto the sphere for comparison.

The model shell shown in Figure 5a has a cubic net cut from a square lattice, which perfectly corresponds to the structure of the 24-mer (3RA0), formed by the Whirly proteins upon binding long DNA molecules and involved in DNA repair mechanisms utilized by plant orgranelles.[45] The eight vertices of the bottom model shell (Figure 5b) are located on the three-fold axes and thus the corresponding positions cannot be occupied by proteins. Leaving these positions vacant yields a structure that perfectly replicates the recently studied chaperon 9NV4. It is formed from small heat shock proteins, which prevent damage that can occur due to protein misfolding in response to elevated temperatures.[46]

Interestingly, the model and observed structures shown in Figure 4 and Figure 5b can be derived from octahedral nets superimposed onto a simple triangular lattice; in the resulting shells, nodes on four-fold axes are eliminated. More details on such nets and their mapping onto a sphere can be found in ref. 47. These and many other protein shells can be also obtained within the framework of the phenomenological theory of spherical crystallization,[48] which is based on the critical density wave approach and the Landau phase transition theory.[49-51]

CONCLUSION

We proposed and thoroughly analyzed a minimalistic model for the sequential nonequilibrium assembly of particles retained on a spherical surface. At each step of the assembly process a single particle is added to the growing shell after which an energy minimization procedure is performed. The shell growth is stopped when adding particles no longer lowers energy of the system. The only parameter of the model is the ratio of the equilibrium distance between a pair of particles interacting via the Lennard-Jones potential to the radius of the sphere.

We have demonstrated that such deterministic assembly process, when each new particle is placed into the position corresponding to the lowest attachment energy, results in many shells with high and intermediate symmetries corresponding to global energy minima for certain numbers of particles. All such high-symmetry and most intermediate-symmetry shells can also effectively assemble via temperature-dependent stochastic pathway, in which positions for the attached particles are chosen using a Boltzmann-type distribution function.

Along with the well-known structures of the simplest icosahedral capsids, our model produces multiple packings with the square-triangular order exhibiting octahedral and tetrahedral symmetries. These model shells correspond to the structures of both synthetic and natural protein assemblies with different functions.

We hope that universality of our results will facilitate the design of protein nanocages, micelles and colloidosomes, architecture of which is pivotal for their functions. Such systems have significant potential for various applications in medicine, materials science and synthetic biology, including delivery or display vehicles,[52] biomedical imaging,[52-53] and biomolecular encapsulation.[51,53]

## AUTHOR INFORMATION

**Corresponding Authors**

*Corresponding author. E-mail: rochal_s@yahoo.fr

**Notes**

The authors declare no competing financial interest.

## ACKNOWLEDGMENT

I.Yu. Golushko, O.V. Konevtsova, D.S. Roshal and S.B. Rochal acknowledge financial support from the Russian Science Foundation, Grant No. 22-12-00105-П.

REFERENCES

(1) Bowick, M. J.; Giomi, L. Two-Dimensional Matter: Order, Curvature and Defects. *Advances in Physics* **2009**, *58*, 449-563. DOI: https://doi.org/10.1080/00018730903043166

(2) Lai, Y. J.; Lin, I. Packings and Defects of Strongly Coupled Two-Dimensional Coulomb Clusters: Numerical Simulation. *Phys. Rev. E* **1999**, *60*, 4743. DOI: https://doi.org/10.1103/PhysRevE.60.4743

(3) Nelson, D. R. *Defects and Geometry in Condensed Matter Physics*; Cambridge University Press: Cambridge, 2002.

(4) Andrade, K.; Guerra S.; Debut, A. Fullerene-Based Symmetry in Hibiscus Rosa-Sinensis Pollen. *PLoS One* **2014**, *9*, e102123. DOI: https://doi.org/10.1371/journal.pone.0102123

(5) Wolken, J. J.; Capenos, J.; Turano, A. Photoreceptor Structures: III. Drosophila Melanogaster. *J. Biophys. Biochem. Cytol.* **1957**, *3*, 441-448. DOI: https://doi.org/10.1083/jcb.3.3.441

(6) Roshal, D. S.; Azzag, K.; Le Goff, E.; Rochal, S. B.; Baghdiguian, S. Crystal-Like Order and Defects in Metazoan Epithelia with Spherical Geometry. *Sci. Rep.* **2020**, *10*, 7652. DOI: https://doi.org/10.1038/s41598-020-64598-w

(7) Dinsmore, A. D.; Hsu, M. F.; Nikolaides, M. G.; Marquez, M.; Bausch A. R.; Weitz, D. A. Colloidosomes: Selectively Permeable Capsules Composed of Colloidal Particles. *Science* **2002**, *298*, 1006-1009. DOI: https://doi.org/10.1126/science.1074868

(8) Bausch, A. R.; Bowick, M. J.; Cacciuto, A.; Dinsmore, A. D.; Hsu, M.F.; Nelson, D. R., Nikolaides, M. G.; Travesset, A.; Weitz, D. A. Grain Boundary Scars and Spherical Crystallography. *Science* **2003**, *299*, 1716-1718. DOI: https://doi.org/10.1126/science.1081160

(9) Flint, S.J.; Racaniello, V.R.; Rall, G.F.; Skalka, A.M.; Enquist, L.W. *Principles of Virology*; ASM Press: Washington DC, USA, 2015.

(10) Caspar D. L. D.; Klug, A. Physical Principles in the Construction of Regular Viruses. *Cold Spring Harb. Symp. Quant. Biol.* **1962**, *27*, 1-24. DOI: https://doi.org/10.1101/sqb.1962.027.001.005

(11) Bruinsma, R. F.; Gelbart, W. M.; Reguera, D.; Rudnick, J.; Zandi, R. Viral Self-Assembly as a Thermodynamic Process. *Phys. Rev. Lett.* **2003**, *90*, 248101. DOI: https://doi.org/10.1103/PhysRevLett.90.248101

(12) Zandi, R.; Reguera, D.; Bruinsma, R. F.; Gelbart W. M.; Rudnick, J. Origin of Icosahedral Symmetry in Viruses. *Proc. Natl. Acad. Sci. U. S. A.* **2004**, *101*, 15556-15560. DOI: https://doi.org/10.1073/pnas.0405844101

(13) Panahandeh, S.; Li, S.; Zandi, R. The Equilibrium Structure of Self-Assembled Protein Nano-Cages. *Nanoscale* **2018**, *10*, 22802-22809. DOI: https://doi.org/10.1039/c8nr07202g

(14) Panahandeh, S.; Li, S.; Dragnea, B.; Zandi, R. Virus Assembly Pathways Inside a Host Cell. *ACS nano* **2022**, *16*, 317-327. DOI: https://doi.org/ 10.1021/acsnano.1c06335

(15) Wagner J.; Zandi, R. The Robust Assembly of Small Symmetric Nanoshells. *Biophys. J.* **2015**, *109*, 956-965. DOI: https://doi.org/10.1016/j.bpj.2015.07.041

(16) Ballard, A. F.; Panter, J. R.; Wales, D. J. The Energy Landscapes of Bidisperse Particle Assemblies on a Sphere. *Soft Matter* **2021**, *17*, 9019-9027. DOI: https://doi.org/10.1039/d1sm01140e

(17) Levandovsky, A.; Zandi, R. Nonequilibirum Assembly, Retroviruses and Conical Structures. *Phys. Rev. Lett.* **2009**, *102*, 198102. DOI: https://doi.org/10.1103/PhysRevLett.102.198102

(18) Paquay, S.; Kusumaatmaja, H.; Wales, D. J.; Zandi, R.; van der Schoot, P. Energetically Favoured Defects in Dense Packings of Particles on Spherical Surfaces. *Soft Matter* **2016**, *12*, 5708-5717. DOI: https://doi.org/10.1039/c6sm00489j

(19) Voogd, J.M. Crystallisation on a Sphere; Computational Studies of Two-dimensional Lennard-Jones Systems. PhD-Thesis University of Amsterdam, June 1998. www.science.uva.nl/research/scs/papers/jeroen.html.

(20) Thomson, J. J. XXIV. On The Structure of the Atom: an Investigation of the Stability and Periods of Oscillation of a Number of Corpuscles Arranged at Equal Intervals Around the Circumference of a Circle; with Application of the Results to the Theory of Atomic Structure. *Philos. Mag.* **1904**, *7*, 237-265.

(21) Berman, H. M.; Westbrook, J.; Feng, Z.; Gilliland, G.; Bhat, T. N.; Weissig, H.; Shindyalov, I. N.; Bourne, P. E. The Protein Data Bank. *Nucleic Acids Res.* **2000**, *28*, 235–242. DOI: https://doi.org/10.1093/nar/28.1.235

(22) Hulo, C.; De Castro, E.; Masson, P.; Bougueleret, L.; Bairoch, A.; Xenarios, I.; Le Mercier, P. ViralZone: a Knowledge Resource to Understand Virus Diversity. *Nucleic Acids Res.* **2011**, *39*, D576–D582. DOI: https://doi.org/10.1093/nar/gkq901

(23) Shepherd, C. M.; Borelli, I. A.; Lander, G.; Natarajan, P.; Siddavanahalli, V.; Bajaj, C.; E. Johnson, J.; Brooks, C. L.; Reddy, V. S. VIPERdb: a Relational Database for Structural Virology. *Nucleic Acids Res.* **2006**, *34*, D386–D389. DOI: https://doi.org/10.1093/nar/gkj032

(24) Widom, M.; Lidmar, J.; Nelson, D. R. Soft Modes near the Buckling Transition of Icosahedral Shells. *Phys. Rev. E* **2007**, *76*, 031911. DOI: https://doi.org/10.1103/PhysRevE.76.031911

(25) Li, S.; Tresset, G.; Zandi, R. From Disorder to Icosahedral Symmetry: How Conformation-Switching Subunits Enable RNA Virus Assembly. *Science Advances* **2025**, *11*, eady7241. DOI: 10.1126/sciadv.ady724

(26) Wingfield, P. T.; Stahl, S. J.; Williams, R. W.; Steven, A. C. Hepatitis Core Antigen Produced in Escherichia coli: Subunit Composition, Conformational Analysis, and In Vitro Capsid Assembly. *Biochemistry* **1995**, *34*, 4919–4932. DOI: https://doi.org/10.1021/bi00015a003

(27) Moerman, P.; Van Der Schoot, P.; Kegel, W. Kinetics versus Thermodynamics in Virus Capsid Polymorphism. *J. Phys. Chem. B* **2016**, *120*, 6003–6009. DOI: https://doi.org/10.1021/acs.jpcb.6b01953

(28) Wales, D.J. A Microscopic Basis for the Global Appearance of Energy Landscapes. *Science* **2001**, *293*, 2067-2070. DOI: 10.1126/science.1062565

(29) Rochal, S. B.; Roshal, A. S.; Konevtsova, O. V., Podgornik, R. Proteinaceous Nanoshells with Quasicrystalline Local Order. *Phys. Rev. X* **2024**, *14*, 031019. DOI: https://doi.org/10.1103/PhysRevX.14.031019

(30) Toussaint, L.; Bertrand, L.; Hue, L.; Crichton, R. R.; Declercq, J. P. High-resolution X-ray Structures of Human Apoferritin H-chain Mutants Correlated with Their Activity and Metal-binding Sites. *J. Mol. Biol.* **2007**, *365*, 440-452. DOI: https://doi.org/10.1016/j.jmb.2006.10.010

(31) Petruk, G.; Monti, D. M.; Ferraro, G.; Pica, A.; D'Elia, L.; Pane, F.; Amoresano, A.; Furrer, J.; Kowalski, K.; Merlino, A. Encapsulation of the Dinuclear Trithiolato-Bridged Arene Ruthenium Complex Diruthenium-1 in an Apoferritin Nanocage: Structure and Cytotoxicity. *ChemMedChem*, **2019**, *14*, 594-602. DOI: https://doi.org/10.1002/cmdc.201800805

(32) Aeschimann, W.; Kammer, S.; Staats, S.; Schneider, P.; Schneider, G.; Rimbach, G.; Cascella, M.; Stocker, A. Engineering of a Functional Gamma-Tocopherol Transfer Protein. *Redox Biol.* **2021**, *38*, 101773. DOI: https://doi.org/10.1016/j.redox.2020.101773

(33) Saibil, H. R. Chaperone Machines in Action. *Curr. Opin. Struct. Biol.* **2008**, *18*, 35–42. DOI: https://doi.org/10.1016/j.sbi.2007.11.006

(34) Kish-Trier E.; Hill, C. P. Structural Biology of the Proteasome. *Annu. Rev. Biophys.* **2013**, *42*, 29–49. DOI: https://doi.org/10.1146/annurev-biophys-083012-130417

(35) Righetto, R.D.; Anton, L.; Adaixo, R.; Jakob, R.P.; Zivanov, J.; Mahi, M.A.; Ringler, P.; Schwede, T.; Maier T.; Stahlberg, H. High-Resolution Cryo-EM Structure of Urease from the Pathogen Yersinia Enterocolitica. *Nat. Commun.* **2020**, *11*, 5101-5101. DOI: https://doi.org/10.1038/s41467-020-18870-2

(36) Maier, T.; Leibundgut, M.; Boehringer, D.; Ban, N. Structure and Function of Eukaryotic Fatty Acid Synthases. *Q. Rev. Biophys.* **2010**, *43*, 373–422. DOI: https://doi.org/10.1017/S0033583510000156

(37) Aeschimann, W.; Staats, S.; Kammer, S.; Olieric, N.; Jeckelmann, J.M.; Fotiadis, D.; Netscher, T.; Rimbach, G.; Cascella, M.; Stocker, A. Self-assembled alpha-Tocopherol Transfer Protein Nanoparticles Promote Vitamin E Delivery Across an Endothelial Barrier. *Sci. Rep.* **2017**, *7*, 4970-4970. DOI: https://doi.org/10.1038/s41598-017-05148-9

(38) Rawson, S.; Bisson, C.; Hurdiss, D.L.; Fazal, A.; McPhillie, M.J., Sedelnikova, S.E., Baker, P.J.; Rice D.W.; Muench, S.P. Elucidating the Structural Basis for Differing Enzyme Inhibitor Potency by Cryo-EM. *Proc. Natl. Acad. Sci. U. S. A*. **2018**, *115*, 1795-1800. DOI: https://doi.org/10.1073/pnas.1708839115

(39) Rochal, S. B., Konevtsova, O. V., Myasnikova, A. E., Lorman, V. L. Hidden Symmetry of Small Spherical Viruses and Organization Principles in "Anomalous" and Double-Shelled Capsid Nanoassemblies. *Nanoscale* **2016**, *8*, 16976-16988. DOI: https://doi.org/10.1039/c6nr04930c

(40) Plevka, P.; Tars, K.; Liljas, L. Crystal Packing of a Bacteriophage MS2 Coat Protein Mutant Corresponds to Octahedral Particles. *Protein Sci.* **2008**, *17*, 1731-1739. DOI: https://doi.org/10.1110/ps.036905.108

(41) Zang, J.; Chen, H.; Zhang, X.; Zhang, C.; Guo, J.; Du, M.; Zhao, G. Disulfide-Mediated Conversion of 8-mer Bowl-Like Protein Architecture into Three Different Nanocages. *Nat. Commun.* **2019**, *10*, 778-778. DOI: https://doi.org/10.1038/s41467-019-08788-9

(42) Peng, P.-P.; Dong, L.-L.; Sun, Y.-F.; Zeng, X.-L.; Ding, W.-L.; Scheer, H.; Yang, X.; Zhao, K.-H. The Structure of Allophycocyanin B from *Synechocystis* PCC 6803 Reveals the Structural Basis for the Extreme Redshift of the Terminal Emitter in Phycobilisomes. *Acta Crystallogr., Sect. D* **2014**, *70* (10), 2558–2569. DOI: https://doi.org/10.1107/S1399004714015776

(43) Dotera, T. Quasicrystals in Soft Matter. *Isr. J. Chem.* **2011**, *51*, 1197-1205. DOI: https://doi.org/10.1002/ijch.201100146

(44) Engel, M.; Trebin, H. R. Self-Assembly of Monatomic Complex Crystals and Quasicrystals with a Double-Well Interaction Potential. *Phys. Rev. Lett.* **2007**, *98*, 225505. DOI: https://doi.org/10.1103/PhysRevLett.98.225505

(45) Cappadocia, L.; Parent, J.S.; Zampini, E.; Lepage, E.; Sygusch, J.; Brisson, N. A. Conserved Lysine Residue of Plant Whirly Proteins Is Necessary for Higher Order Protein Assembly and Protection Against DNA Damage. *Nucleic Acids Res*. **2012**, *40*, 258-269. DOI: https://doi.org/10.1093/nar/gkr740

(46) Miller, A.P.; Reichow, S.L. Mechanism of Small Heat Shock Protein Client Sequestration and Induced Polydispersity. *Nat. Commun.* **2025**, *16*, 3635. DOI: https://doi.org/10.1038/s41467-025-58964-3

(47) Chalin, D. V.; Rochal, S. B. Landau theory and Self-Assembly of Spherical Nanoclusters and Nanoparticles with Octahedral Symmetry. *Phys. Rev. B*. **2023**, *107*, 024102. DOI: https://doi.org/10.1103/PhysRevB.107.024102

(48) Konevtsova, O.V; Chalin, D. V.; Roshal, A.S.; Rochal, S. B. Non-Viral Protein Shells with Octahedral and Tetrahedral Symmetries and Theory of Critical Density Waves, Submitted to *Journal of the Royal Society Interface* **2025**.

(49) Lorman, V. L.; Rochal, S. B. Density-Wave Theory of the Capsid Structure of Small Icosahedral Viruses. *Phys. Rev. Lett.* **2007**, *98*, 185502. DOI: https://doi.org/10.1103/PhysRevLett.98.185502

(50) Landau, L. D. On the Theory of Phase Transitions. *Zh. Eksp. Teor. Fiz.* **1937**, *11*, 19-32.

(51) Landau, L. D. On the Theory of Phase Transitions. II. *Zh. Eksp. Teor. Fiz.* **1937**, *11*, 627.

(52) Schoonen, L.; van Hest, J.C.M. Functionalization of Protein-Based Nanocages for Drug Delivery Applications. *Nanoscale* **2014**, *6*, 7124–7141. DOI: https://doi.org/10.1039/c4nr00915k

(53) Aumiller, W.M.; Uchida M.; Douglas, T. Protein cage assembly across multiple length scales. *Chem. Soc. Rev.* **2018**, *47*, 3433–3469. DOI: https://doi.org/10.1039/c7cs00818j